\documentclass[prb,preprint,showpacs,preprintnumbers,amsmath,amssymb]{revtex4}

\usepackage{graphicx}
\usepackage{dcolumn}
\usepackage{bm}

\begin{document}


\title{Formation of Pt islets on facets of Ru nanoparticles: a first principles study}

\author{Marisol Alc{\'a}ntara Ortigoza}
\email{alcantar@physics.ucf.edu}
\author{Sergey Stolbov}
\email{sstolbov@physics.ucf.edu}
\author{Talat S. Rahman}
\email{talat@physics.ucf.edu}
\affiliation{Department of Physics, University of Central Florida\\
Orlando, Florida 32816, USA
}%

\date{\today}

\begin{abstract}
We have carried out density functional theory based calculations of the size dependent formation energy and geometry of Pt islands on Ru(0001), to model Pt-Ru nanocatalysts which have been recently proposed as fuel cell anode. The Pt islands are found to prefer two-dimensional structures. Furthermore, a monotonic decrease in the formation energy per Pt atom suggests a propensity of Pt atoms to wet the Ru(0001) surface. Calculated energy barriers for the diffusion of Pt monomers and dimers on the facets and through the edges of a superstructure modeling a Ru nanoparticle indicate that these edges help reduce considerably the diffusion rates across them such that the Pt atoms prefer to remain in the facet on which they were adsorbed originally and form 2D islands.

\end{abstract}

\maketitle

\section{Introduction}\label{prI}

Direct methanol fuel cells (DMFC) are considered a promising means for energy conversion in "hydrogen-based economy" because they work at low temperatures ($\sim$~350-400 K) and use liquid methanol as fuel, which is easy to deliver and store.
In DMFC, the same anode is used as a catalyst for both methanol reforming and for the oxidation of hydrogen obtained from the reforming. Although the carbon monoxide released in the course of this reaction is expected to be oxidized by hydroxyl radicals obtained from admixed water, experiments find that it still severely poisons the commonly used Pt anode by blocking the reactive Pt sites and, thereby, reducing the rate of hydrogen electro-oxidation. Similarly, in proton exchange fuel cells operating with pre-reformed gas, the anode is poisoned by carbon monoxide molecules, inevitably present in hydrogen obtained from hydrocarbons.

It is known~\cite{p1} that PtRu alloys are more tolerant to CO poisoning than pure Pt, though their tolerance is still unsatisfactory. The high content of expensive platinum in these alloys also speaks against their choice as cost effective catalysts. It is thus encouraging to note that nanoclusters of Ru exposed to spontaneous Pt deposition are much more tolerant to CO than commercial PtRu catalysts,~\cite{p2,p3} particularly since the content of Pt on these novel materials is significantly smaller than that in Ru-Pt alloys. For example, Brankovic \emph{et al}.~\cite{p2,p3} deposited Pt
on $\sim$2.5 nm size Ru nanoparticles and found that the 1:20 ratio (PtRu$_{20}$), which corresponds to $\sim$~0.1 monolayer (ML) coverage, surpass substantially the catalytic performance of PtRu and Pt$_2$Ru$_3$ in
the presence of CO. Assuming that Pt atoms form two dimensional (2D) islands on the facets of the Ru nanoparticles,~\cite{p3} the CO tolerance was attributed~\cite{p3} and explained~\cite{sergey} on the basis of the \emph{spillover} effect, which refers to the tendency of CO to leave the Pt island and diffuse towards Ru.

As noted above, Pt coverage is critical for the catalytic properties of the Ru nanoparticles.~\cite{p2,uribe} Island size effects - tuned by Pt coverage - have also been reported for methanol electro-oxidation.~\cite{p4} Earlier calculations have indicated that the CO adsorption energies on binary Pt-Ru systems, ranging from Pt(111) to surfaces of ordered Pt-Ru alloys to 1ML of Pt on Ru(0001), are reduced upon lowering Pt content.~\cite{p5} Moreover, our recent calculation~\cite{sergey} supporting the idea of CO spillover relies on the presence of small 2D Pt islands on the facets of Ru nanoparticles. All of the above suggest that the correlation between catalytic activity and coverage has implications on how Pt arranges itself on the Ru nanoparticles, as we shall see later. Indeed, the first step in obtaining a systematic understanding of the enhanced reactivity of this catalyst is the determination of the geometry and relative stability of the Pt islands on the nanoparticle facets. Towards this end we have adopted in this work a theoretical approach based on \emph{ab initio} electronic structure calculations.

Ever since the work by Stranski in 1927 about crystal growth, a large amount of effort has been dedicated to understanding the microscopic processes that control nucleation and epitaxial
growth.~\cite{markov} It is now recognized that the resulting
structure when adatoms are adsorbed onto a surface is thermodynamically
and kinetically controlled. The energetic barriers for the
various processes to occur, the surface and/or post-growth annealing temperature, along with the deposition rate and technique, essentially determine the morphology of the deposited adatom structure at a given coverage. Three major types of growth modes have been observed in experiment.~\cite{chang} The Frank van der Merwe  mode refers to the two dimensional (2D) layer-by-layer growth, the Volmer-Weber growth describes three-dimensional (3D) clustering of adatoms on the bare substrate, while the Stranski-Krastanov mode consists of a mixture of the above two growth modes, i.e., 3D clustering occurs after one or a few complete adlayers are formed.~\cite{chang, markov}
Theoretical models of heteroepitaxial growth suggest that the growth mode is determined by the competition of factors such as the surface energies of the bare substrate and the heteroepitaxial
layer, the interface free energy, and the strain energy introduced by the lattice mismatch of the two species.~\cite{chang, markov, copel} For example, a mismatch in the respective bulk
lattice constants strains the interface and may set off the 3D clustering growth mode.~\cite{chang}
On the contrary, growth of adlayers with lower surface energy than the substrate may favor the 2D layer-by-layer growth mode.~\cite{chang}
Considerations along these lines, however, lead to ambiguities in predicting Pt growth on Ru(0001).
While the higher cohesive energy of Ru (relative to Pt) may~\cite{shutter, kummi} imply
that the surface free energy of Ru(0001) is  higher than that of Pt(111)
and point to 2D layer-by-layer growth, the
 stress caused by the Ru-Ru, Ru-Pt, and Pt-Pt bond length misfit (The bond lengths in bulk Ru and Pt are 2.706~{\AA} and 2.775~{\AA}, respectively.~\cite{kittel}) may lead to 3D clustering.
 Of course, there may be a competition between the above two factors, leading to a critical Pt island size at which there is crossover between 2D and 3D growth mode or island-substrate atom exchange.~\cite{p6,p7}
 One of the goals in this work is to determine whether there is indeed a critical size beyond which 2D Pt islands are no longer stable on Ru(0001).

Turning to the case of the Ru nanoparticles described in the experiments in question,~\cite{p2,p3} it is important to note that, while the experimental determination of the structure and chemical ordering of Pt-Ru nanoalloys represented a challenge\cite{russell,mcbreen,hills} until the recent work of Maillard \emph{et al}.,\cite{maillard} the dominant features observed via surface X-ray scattering, scanning tunneling microscopy,
Fourier transform infrared spectroscopy, and high resolution transmission electron microscopy techniques, in 2-3 nm pure Ru nanoparticles on a carbon substrate have been consistent with the hexagonally
close-packed (hcp) Ru single crystal structure.\cite{p3,maillard}
The Ru nanoparticles present well-defined facets with the lowest-surface-energy geometries, such as (0001), ($\overline{1}$101), etc.\cite{p3,maillard} In the present study, as a
first step in the modeling of the Pt-decorated Ru nanoparticles, we focus our attention on the
formation of Pt islands as a function of size
on Ru(0001), one of the most predominant and stable facet orientations.\cite{p19,p3,maillard} To this end, we carry
out first principles calculations of the system's total energy to determine the geometry and formation energy of Pt islands, as well as that of 1ML of Pt on Ru(0001).
Clearly, the main drawback of such calculations is
that the predictions are relevant to zero temperature and samples
relaxed in \emph{infinite time}. To partially include the system dynamics, we have taken
into account the  diffusivity of Pt adatoms since, particularly at temperatures well
below room temperature, the three aforementioned growth modes can be
understood in terms of diffusion barrier differences between on-step and step-descending hopping, Schwoebel barriers.~\cite{schoebel, tsong1, tsong2}
It is argued that if the energy barrier at the step edge is higher (reflective barrier)
than that of adatom diffusion
 on the step - positive Schwoebel barrier - there is a probability that adatoms will be trapped on the step terraces and 3D clustering will be favored.
On the contrary, if the Schwoebel barrier is zero or negative (non-reflective barrier), adatoms that happen to lie on
the step terrace are more likely to hop onto the substrate and favor the 2D growth.~\cite{tsong1}
Moreover, in heteroepitaxial growth, the mixed Stranki-Krastanov mode dominates if the Schwoebel barrier
is reflective for homo-steps (formed by the adatoms), but non-reflective for hetero-steps.~\cite{tsong1, tsong2}

Furthermore, since Ru nanoparticles present facets with various geometries, it is reasonable to assume that the equilibrium shape is a polyhedron and, therefore, there are edges delimiting the facets. Theoretical\cite{kinoshita1,kinoshita2} and experimental\cite{amalija} studies of the shape of Pt nanoparticles supported on carbon, for instance, have consistently concluded that they are in the form of cubo-octahedra, which is a shape characteristic of fcc packing. For hcp metals, in turn, one of the proposed structures for sufficiently large nanoparticles (with more than 500 atoms) is the so-called anticubo-octahedron\cite{ryan} or the truncated hexagonal bipyramid.\cite{barcaro2} The essential point is that, since the above polyhedra have several facets (14 for the cubo- and anticubo-octahedron\cite{ryan} and 20 for the truncated hexagonal bipyramid\cite{barcaro2}), one expects the morphology of the adsorbed Pt to be sensibly influenced by the diffusivity of Pt adatoms/islands through the edges of the Ru nanoparticles. To assess the possibility of diffusion of Pt monomers and dimers, we simulate two edges of a Ru nanoparticle using a superstructure described in the next section and calculate diffusion barriers of Pt monomers and dimers on its (0001) and ($\overline{1}$101) facets and across its edges. Note that our model for the facet geometries coincide with some of those detected in experiment\cite{maillard,p3} and are also proper to the truncated hexagonal bipyramid.\cite{barcaro2}

The rest of this article is organized as follows: Section~\ref{prII} presents the computational details, Section~\ref{prIII} contains our results and provides some discussion about Pt islands on Ru(0001) (subsection~\ref{prIIIA}) and of Pt diffusion on the (0001) facet and across the edges of our Ru nanostructure (subsection~\ref{prIIIB}). Section~\ref{prIV} summarizes our results and conclusions.

\section{Computational details}\label{prII}

Periodic supercell calculations have been carried out within the density functional theory (DFT) framework,~\cite{p10,p11} using the plane wave and pseudopotential methods~\cite{p12} as embodied into VASP (Vienna \emph{ab initio} Simulation package),~\cite{p13} and with ultrasoft pseudopotentials.~\cite{p14} We have used a kinetic energy cutoff of 400 eV for the wave functions and 700 eV for the charge density to obtain convergent results with
sufficient computational accuracy of the lattice constant of bulk Ru and Pt. Brillouin zones were sampled with either the $(4 \times 3 \times 1)$ or the $(3\times 3\times 1)$ Monkhorst-Pack k-point meshes,~\cite{p15} depending on the size of the supercell, as we will see. Since the main uncertainty of DFT comes from the exchange-correlation potential, we have used two different approximations for the exchange-correlation functional: the Perdew and Wang generalized gradient approximation (GGA)~\cite{p16} and the Perdew, Burke, and Ernzerhof (PBE) modified GGA~\cite{p17}, and compared some of the results obtained using these two approximations. To achieve force relaxation of the studied structures, the total energy of the system and the forces acting on each atom are obtained after each self-consistent electronic structure and minimized by the conjugated-gradient algorithm.~\cite{p18} At equilibrium, forces on each atom are required to be below 0.02 eV/~{\AA}. The diffusion barriers for monomers and dimers on the Ru superstructure are obtained by the direct dragging method. The 3D graphics presented in this work were generated by the \emph{Xcrysden} program.~\cite{xcrys}

We note that in the calculations of the geometry and formation energy of Pt islands the usage of supercells may introduce contributions from island-island interactions, as a result of the imposed periodicity of the system. This is particularly true for the largest islands in the $3\times 4$ and $4\times 4$ supercells in which edge-atoms of neighboring islands are as close as third nearest neighbors (NN) of the Ru(0001) surface.
A simple way to estimate this spurious interaction is to consider the interaction energy, $E^{int}$, between two Pt atoms adsorbed on Ru(0001) as their separation varies and defined as $E^{int} = E(2$Pt$/$Ru$) - 2E(1$Pt$/$Ru$)$, where  $E(2$Pt$/$Ru$)$ is the adsorption energy of 2 Pt atoms on Ru(0001) in a $4 \times 4$ supercell and $E(1$Pt$/$Ru$)$ is the adsorption energy of 1 Pt atom on the same surface and supercell. We find that $E^{int} =$ -0.203, +0.074, and -0.018 eV, as the separation between the two Pt atoms increases from $1^{st}$ to $2^{nd}$, and $3^{rd}$ NN bond length, respectively. The interaction between $3^{rd}$ NN is thus expected to be 10 times smaller than that of $1^{st}$  NN. Accordingly, $E_{form}$, as calculated in this work, can be reliable up to $\pm$ 0.02 eV for the largest islands.

\subsection{2D Pt islands on Ru(0001)}\label{prIIA}

The facets of Ru nanoparticles are first modeled by a 5 layer Ru(0001) slab, which is the  surface known to have the lowest energy.~\cite{p19} Pt adatoms are placed on only one side of the slab. To avoid the interaction between surfaces and Pt adatoms of neighboring periodic supercells we have imposed a 15~{\AA} vacuum layer, whereas to reduce the interaction between deposited
Pt islands, the (0001) surface unit cells is extended to either $(3\times 4)$ or $(4\times 4)$ structures depending on the island size. The $(3\times 4)$  and $(4\times 4)$ supercells contained 60 and 80 Ru atoms, respectively, plus Pt atoms forming the island. Their corresponding surface Brillouin zone is sampled with a $(4\times 4 \times 1)$  and $(3\times 4 \times 1)$  k-point mesh, respectively.

\subsection{Monomer and dimer on faceted superstructure}\label{prIIB}

\begin{figure}
\includegraphics[width=0.53\textwidth]{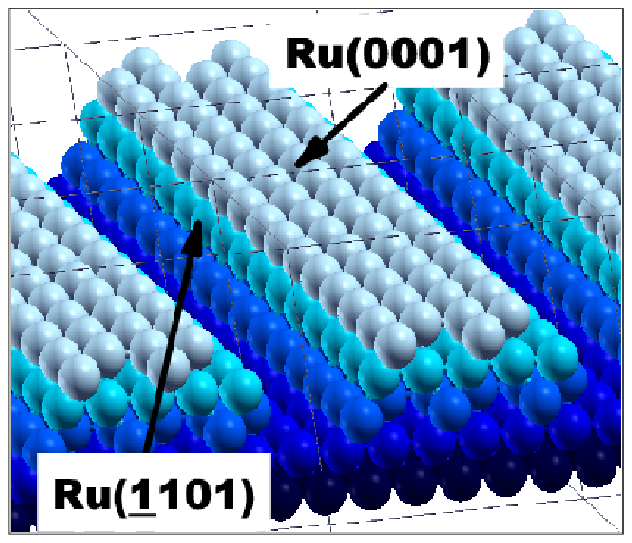}
\caption{\label{fig:pr1}}
\end{figure}

To model Pt diffusion through the Ru nanoparticles edges we have taken into consideration edges formed by facets of (0001) and ($\overline{1}$101) geometry, which are among the most stable Ru surfaces.~\cite{p3,p19} We consider a periodic 3D superstructure containing 116 Ru atoms and made of a 4-atom wide Ru(0001) facet and two Ru($\overline{1}101$) facets (see Fig.~\ref{fig:pr1}).
The construction of this Ru supercell, which has $7 \times 4$ in-plane periodicity, is achieved by stacking five Ru(0001) layers:  two of $7 \times 4$, one of $6 \times 4$, one of  $5 \times 4$, and one of $4 \times 4$ atoms. The so obtained edges, on each side of the Ru(0001) have different local geometry which for convenience are labeled as $A$ and $B$. Atoms forming edge $A$ (edge $B$) are contiguous to hcp (fcc) hollow sites of the (0001) facet. The bottom two layers (see Fig.~\ref{fig:pr1}) were not allowed to relax to guarantee the stability of the superstructure. We impose a 15~{\AA} vacuum layer between
periodic superstructures along the direction perpendicular to the surface, as in the system described previously. The Brillouin zone is sampled with a $(2 \times  3 \times 1)$ k-point mesh. The adsorption energy and diffusion barriers of Pt monomers and dimers are calculated on the (0001) and the ($\overline{1}$101) facets.

\section{Results and Discussion}\label{prIII}

To determine in-plane slab periodicity, we have calculated bulk lattice parameters using PW and PBE approximation for the exchange-correlation functional. The bond lengths of bulk Ru and Pt are found to be 2.706(2.744PBE) and 2.77(2.82PBE)~{\AA}, respectively, while the c/a ratio of bulk Ru is found to be 1.585 (PBE)~{\AA}.

\subsection{Pt islands on Ru(0001)}\label{prIIIA}

We have calculated the optimized geometric structure and energetics of 1 to 5 Pt atom islands adsorbed on Ru(0001) using the $(3 \times 4)$ supercell and that of 1 to 9 Pt atom island on Ru(0001) using the $(4 \times 4)$ supercell. The relaxed structure and total energy of one Pt monolayer on Ru(0001) has been also obtained. To characterize the energetic stability of a given Pt island, we obtain its formation energy, which is defined as: $E_{form} = E($Ru$+$Pt$) - E($Ru$) -  nE($Pt$_{at})$. Here, $E($Ru$+$Pt$)$ is the total energy of a Ru slab adsorbed with a $n$-atom Pt island, while $E($Ru$)$ and  $E($Pt$_{at})$ denote the total energies of the clean Ru slab and a free Pt atom, respectively. Note that the formation energy of stable structures is negative. The structure with lowest average formation energy per Pt atom, $E_{form}/n$, will thus be distinguished as the energetically most favorable one.

\begin{figure}
\includegraphics[width=0.53\textwidth]{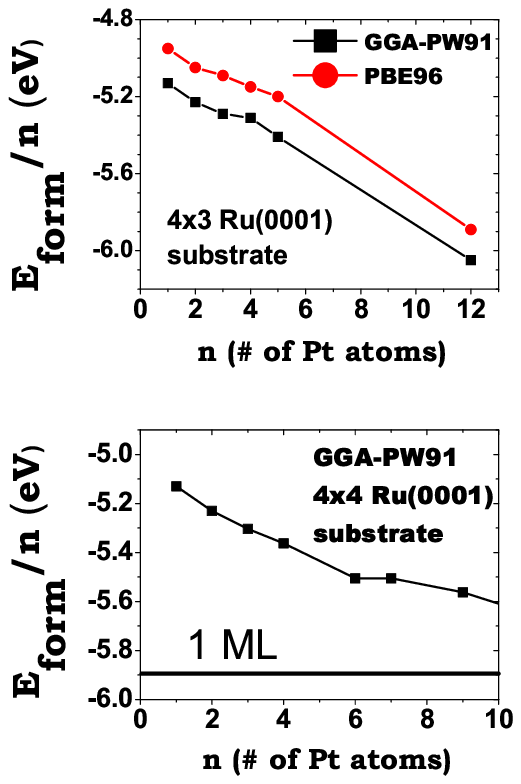}
\caption{\label{fig:pr2}}
\end{figure}

Fig.~\ref{fig:pr2}(a) presents $E_{form}$/n as a function of
the size of the island, $n$, for $n =$ 1-5 on the $(3 \times 4)$ supercell,
as provided by PBE and GGA. Fig.~\ref{fig:pr2}(b) displays the
dependencies of $E_{form}/n$,
for $n =$ 1-4, 6, 7 and 9, on the $(4 \times 4)$ supercell. Note that the values $n =$ 12 and $n =$ 16
represent full monolayer coverage of Pt for the $(3 \times 4)$ and $(4 \times 4)$ supercells,
respectively. We find that both PBE and GGA provide similar qualitative dependencies: the
larger island, the stronger the bonds. Such a trend culminates and
is confirmed at full monolayer Pt coverage, which provides lowest $E_{form}/n$.

\begin{figure}
\includegraphics[width=0.5\textwidth]{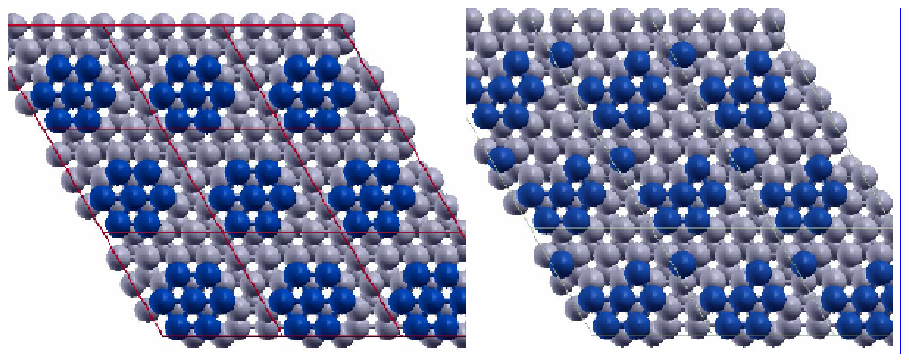}
\caption{\label{fig:pr3}}
\end{figure}

The effect of atom detachment (from Pt islands) on $E_{form}$ has been studied as well.
Fig.~\ref{fig:pr3} shows two configurations considered for
the 7-atom Pt island adsorbed on Ru(0001). We find that the
detachment (transition from the left to the right configuration
in the figure) causes an increase in $E_{form}$
from -38.55 eV to -37.52 eV. Similar results have been
obtained for the islands of other sizes.
The increment of energy per Pt atom upon detachment, however,
does not depend significantly on the island size
and vary in the range of 0.11 - 0.14 eV.  For instance,
for a 2-atom Pt island, detachment leads to an increase
in $E_{form}$ from -10.46 eV to -10.24 eV while, for a 3-atom island,
the energy increases from -15.88 eV to -15.52 eV upon the detachment.

\begin{figure}
\includegraphics[width=0.5\textwidth]{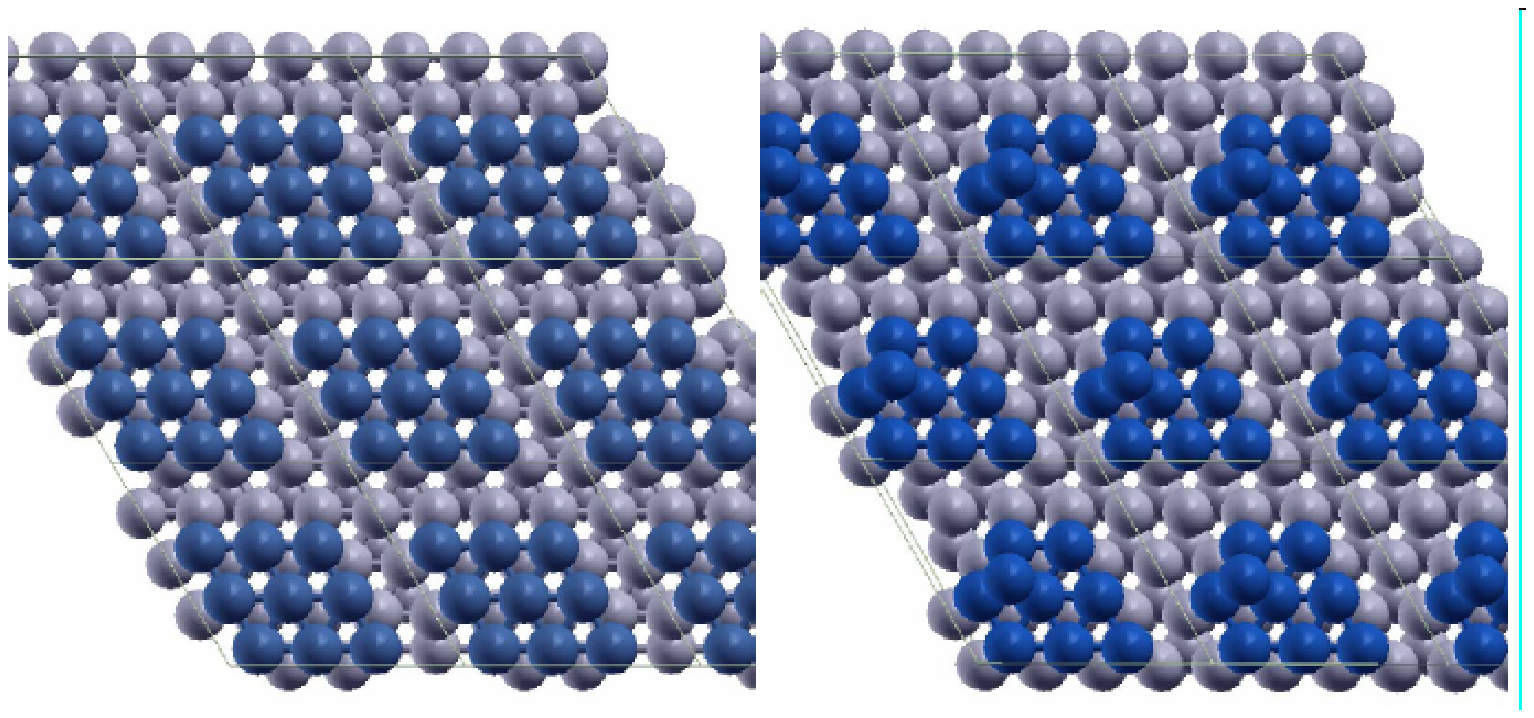}
\caption{\label{fig:pr4}}
\end{figure}

We thus obtain a clear trend: the larger a two-dimensional Pt island (up to 1 ML) is, the lower its formation energy per atom. Thus, assuming that the free energy of the system in consideration is dominated by its DFT total energy, we conclude that Pt tends to wet Ru(0001).
We have also performed calculations for some 3D Pt islands on
Ru(0001) and found that their 2D isomers have lower energy. For example, two configurations of a 9-atom Pt island with 2D and 3D structures (see Fig.~\ref{fig:pr4}) were found to have $E_{form} =$ -50.06 eV and -48.54 eV, respectively.


The above results indicate that Pt atoms overpower the stress effects derived from the Pt-Ru bond length misfit and have a propensity to increase their local coordination on Ru(0001) regardless of the chemical nature of the bonds, thus forming 2D islands up to $\sim$~0.56 ML and, presumably, up to 1 ML (see Fig.~\ref{fig:pr2}). Of course our calculations do not include the case of larger Pt islands (more than 10 atoms) and leave open the question of formation of 3D structures in such cases. Nevertheless, there is experimental evidence that Pt may form 2D pseudomorphic islands up to full coverage through vapor deposition.\cite{p8}
In turn, from the standpoint of \emph{ab initio} calculations, our conclusion that Pt grows pseudomorphologically is justified only at low temperatures and after a long relaxation time. To take into account the kinetics of the system, we have looked at the diffusivity of a Pt adatom in the event that it be adsorbed on top of Pt islands. Consider, for example, the case presented in Fig.~\ref{fig:pr4}. The question is thus whether the Pt adatom will be trapped on top of the Pt island (as shown in the right inside of Fig.~\ref{fig:pr4}) or will it descend at the kink site on the Ru(0001) surface (left inside of Fig.~\ref{fig:pr4}).
Our calculated activation barrier of the Pt monomer diffusion on the 8-atom Pt island from the hcp to fcc sites is~$\sim$~0.23 eV
and from the fcc to hcp is $\sim$ 0.09 eV. The activation barrier for the step descent for a Pt monomer from the fcc to the kink site on the Ru(0001) is $\sim$~0.39 eV (and $\sim$~1.92 eV for the inverse process). Therefore, the Schwoebel barrier of this heteroepitaxial step descent of Pt monomers is $\sim$ +0.30 eV, pointing to 3D clustering growth.
These opposing results hint why the growth mode of Pt on Ru(0001) sensibly depends on experimental conditions.\cite{kasberger,p3} Spontaneous deposition of Pt, for example, has led to the conclusion that 3D structures follow the 2D pseudomorphic islands if the Ru nanoparticles are immersed for sufficiently long periods of time (not specified) into a 0.1M HClO$_4$ solution containing 0.1mM H$_2$PtCl$_6$,\cite{p27} whereas spontaneous deposition by two-minute immersion into 0.1 mM [PtCl$_6$]$^2-$ and subsequent rinsing with 0.1 M H$_2$SO$_4$ and ultra pure water lead to Pt nanoparticles with columnar shape of 10-15 MLs and 3-5 nm.\cite{p9}
For the purpose of our analysis, however, we shall notice that both vapor deposition below 0.6 ML\cite{kasberger} and spontaneous deposition via \emph{short} immersion into Pt-containing solutions\cite{p27} find that the probability for Pt  adatom to arrive on top of a Pt island is low. Indeed, the Ru nanoparticles that we are trying to model are decorated with Pt at very low coverage ($\sim$ 0.1 ML) via spontaneous deposition by immersion in a [PtCl$_6$]$^{2-}$ and rinsed with 0.1 M H$_2$SO$_4$ solution at room temperature,~\cite{p2} therefore, only formation of 2D islands is expected. Besides, there is indication that, even in the case of positive Schwoebel barriers, as long as the atom deposition rate is low and the diffusion speed fast, the reflective property of steps is valid only at temperatures well below room temperature for fcc(111) metals.~\cite{tsong2}

There still remains the question: why, despite the misfit, 2D configurations are favorable? In order to grasp further understanding of the issue, one must realize that a number of counter-trends must be taken into account. Firstly, the 2D growth is not all that surprising for the reason that the bond-length misfit between Pt and Ru amounts only to a few percent. Secondly, in the 2D Pt islands under consideration, the number of NN fluctuates from 3 (single atom) to 9 (full monolayer). Decrease in the number of NN usually causes reduction of the equilibrium interatomic distances. Indeed, we find that for a free standing Pt monolayer, in which every Pt atom has only 6 nearest neighbors, the equilibrium Pt-Pt NN distance is much shorter ($\sim$2.6~{\AA}) than that in bulk Pt ($\sim$2.8~{\AA}) and even shorter
than the Ru-Ru NN distance in bulk Ru ($\sim$2.71~{\AA}). The misfit in low dimensional structures is thus not a well defined quantity because of the dependency on the coordination number of the atoms in question. Correspondingly, the issue of stress induced by the bond-length misfit between the Pt nanostructures and the Ru surface atoms, neither one of which is expected to be at the bulk value given the diversity of their local geometric environment, is thus not clear for these heteroepitaxial system.

In addition, Barcaro \emph{et al}. have pointed out that heteroepitaxial Volmer-Weber (3D) growth is favored for those systems in which the adsorbed metal interacts more weakly with the substrate than with atoms of the same species.\cite{barcaro} In our particular case, however, we obtain that the Ru-Pt bond may in fact be stronger than that of Pt-Pt since our calculations show that the average formation energy per atom of a Pt monolayer on Ru(0001) ($\sim$~5.89 eV) is higher than both the cohesive energy of Pt bulk ($\sim$~5.85 eV) and the average cohesive energy per atom of the Pt surface ($\sim$~5.6 eV).

For the Pt atoms on the top of a hcp metal such as Ru, there is also an incommensurability in bulk structure.
We find that the bulk NN bond length of Pt atoms certainly decreases when they arrange in an hcp structure. In that case, the bulk bond length misfit between Pt and Ru decreases from $\sim$~2.8 to $\sim$~1.4~{\%}. Furthermore, the surface \emph{interlayer} distance in Pt(111) expands to 2.49~{\AA} ($\sim$~1.0~{\%} with respect to bulk), while that of the hypothetical Pt(0001) contracts to 2.39~{\AA} notwithstanding that \emph{intralayer} NN distances are 1.8~{\%} smaller than in the fcc bulk. Therefore, both the reduced coordination in adsorbed Pt clusters and the Pt-Ru bond strength may quench the anticipated effect of bulk bond-length misfit in such a way that Pt atoms adsorbed on Ru(0001) surface tend to form 2D islands as large as possible up to one monolayer.


\subsection{Modeling Pt diffusion on the (0001) facets and through the edges of Ru nanostructures}\label{prIIIB}

From the previous section, we have gained an understanding of the tendency of Pt atoms to form 2D islands,  wetting the Ru(0001) surface rather than clustering in multiple 2D or 3D structures. The experimental evidence, however, suggests that Pt
islands maintain small sizes on Ru nanoparticles,~\cite{p3} say $\sim$0.5 nm (5 to 7 atoms), for  0.1 ML coverage. The difference in the characteristic of Pt surface alloys on Ru(0001) and on Ru nanoparticle is  part of the reason for the substantial reactivity of Pt adatoms on Ru nanoclusters and not on Ru surfaces. If Pt atoms were to make as many bonds as possible on the Ru nanoparticle as they do on the Ru surface, one would expect that, even for low coverage ($\sim$0.1 ML), a large island should totally cover one of the facets of the Ru nanoparticles. For example, a hcp Ru nanoparticle of 2.5 nm with the proposed anticubo-octahedral structure~\cite{p3,ryan} could hold roughly $\sim$7 ($\sim$4) Pt atoms per squared (triangular) facet for homogeneous coverage of 0.1 ML ($\sim$~70 Pt atoms), whereas the same coverage coalesced into a single island could totally cover one of the squared facets, which seems not to be the case.~\cite{p3} Note that a similar argument would follow for a truncated hexagonal bipyramid.

One main difference between the \emph{infinite} surface and the nanoparticle is that the latter exhibits edges dividing its facets. It is natural to assume that they prevent Pt coalescence on the Ru nanoparticles. If this is true, 2D islands are formed on each facet, but they do not join together into a large unique island because the edges prevent those initial small islands from diffusing to other facets, thus persisting as few-atom 2D islands. Support for the aforesaid reasoning will be attained below by comparing the barrier for Pt atoms to diffuse on a (0001) facet with that for Pt to diffuse across the edges towards a ($\overline{1}$101) facet.

\subsubsection{Pt Monomers}\label{prIIIB1}

The Ru nanostructure used for the above purpose is displayed in Fig.~\ref{fig:pr1}. It possesses 3 hcp (hcp1, hcp2, hcp3) and 3 fcc (fcc4, fcc5, fcc6) non-equivalent hollow sites on the (0001) facet, as shown in Fig.~\ref{fig:pr5}(a). The calculated adsorption energies ($E_{ads}$) of Pt monomers on these sites are listed in Table~\ref{tab:tpr1}. We find that for all hcp sites $E_{ads}$ is higher than for any fcc site. Table~\ref{tab:tpr1} shows in addition that Pt monomers preferably sit on sites surrounded by 2 edge atoms (fcc4 and hcp3), rather than on those surrounded by only one (fcc6 and hcp1) or none (fcc5, hcp2) edge atom. Note also that the adsorption energy of a monomer on the (0001) facet is lower than on the infinite surface (5.13 eV). Across edge $A$, the first ($\overline{1}$101)  available site is a 4-fold hollow site, denoted by "1" in Fig.~\ref{fig:pr5}(b), whose adsorption energy, 5.66 eV, is substantially higher than that on hcp sites of the (0001) facet. Across edge $B$, the first ($\overline{1}$101) available site is a 3-fold hollow site, denoted by "2" in Fig.~\ref{fig:pr5}(b), whose adsorption energy is 4.92 eV. The above results, incidentally, suggest that there may be a propensity of Pt to deposit on ($\overline{1}$101) facets in the long range.

The calculated diffusion barriers, $\Delta E$, through edge $A$, appear to be highly asymmetric (see Fig.~\ref{fig:pr6}(a)): $\Delta E$(hcp3 $\rightarrow$ "1") = 0.49 eV and $\Delta E$("1" $\rightarrow$ hcp3) = 1.10 eV. On the other hand, the shortest path for diffusion through the edge B connects the hcp1 site on the (0001) facet with the three-fold hollow site on the ($\overline{1}$101) facet (fcc4 $\rightarrow$ "2"). The barrier for diffusion along this path is found to be 0.22 eV, which is just slightly higher than the barriers for diffusion on the facet. However, our calculations indicate that the initial state for this process (hcp1 site) is metastable since its total energy is 80 meV higher than that for the monomer adsorbed on the neighboring fcc4 site. The diffusion of the Pt monomer through the edge B is thus a two-step process with diffusion from the hcp1 to fcc4 site on (0001) and then from that to the tree-fold site on ($\overline{1}$101). For the above reason, the overall probability of diffusion through the edge B is substantially reduced. Even at room temperature, the probability to find a Pt adatom on the fcc4 site is $\sim$~24 times lower than that on the hcp1 site. Moreover, the hcp1~$\leftrightarrow$~fcc4 barrier is asymmetric (0.20 eV for hcp1~$\rightarrow$~fcc4 and 0.12 eV to diffuse back), which further reduces the diffusion probability.

In order to estimate the probability of Pt diffusion predicted by these barriers, we have calculated roughly the diffusion rate of Pt monomers.
The latter, dominated by the exponential of the energy barrier, is given by $R = D_0 e^{- \frac{\Delta E}{k_BT}}$, where $k_B$ is the Boltzmann constant, $T$ is the temperature, and $D_0$ is the diffusion prefactor~\cite{p26} whose typical values are of the order of $\sim$ 10$^{12} s^{-1}$. By noting from above that the diffusion rate through the edge $B$ is reduced by a factor of $\sim$~24 with respect to that from fcc4~$\rightarrow$~"2", we obtain that the former is effectively $\sim$~6$\times$10$^6~s^-1$, thus significantly smaller than that on the (0001) facet ($\sim$ 4$\times$10$^8$ and $\sim$1$\times$10$^{10}  s^{-1}$, for hcp1 $\rightarrow$  fcc4 and fcc4 $\rightarrow$  hcp1, respectively). On the other hand, the diffusion rate through edge A ($\sim$~3$\times$10$^{-7}$ and $\sim$~6$\times$10$^3~s^-1$, for "1"~$\rightarrow$~hcp3 and hcp3~$\rightarrow$~"1", respectively) is found to be three orders of magnitude lower than that through edge $B$ and at least five orders of magnitude lower than that on the (0001) facet. Notice that  since the barriers to diffuse back from the ($\overline{1}$101) facet to the (0001) facet through edge $B$ ("2"~$\rightarrow$~fcc4) is only slightly larger (0.28 eV) than the barrier for diffusion in the opposite direction, diffusion across edge  B may be actually inefficient. The rate at which Pt monomers return to the (0001) facet can nevertheless be expected to be lower than that by which they diffuse to the neighboring four-fold hollow site on the ($\overline{1}$101) facet, since the barriers for the latter process are significantly lower. The diffusion from the three-fold hollow site to the four-fold hollow site on the ($\overline{1}$101) facet is found to be a two-step process. It involves the diffusion to an intermediate local minimum posing a 0.12 eV barrier and a subsequent step with a barrier of only 70 meV. The barriers to return to the three-fold hollow site are, correspondingly, 0.20 eV and 0.6 eV. These values confirm that, at least energetically, monomers may have a propensity to remain on the ($\overline{1}$101) facet once they occupy it. However, exact implications of these rates and the impact of competing processes can only be visualized after kinetic effects are properly included, such as in kinetic Monte Carlo simulations which we leave as a further task. 
In summary, our results indicate that edges compel Pt monomers to remain on the facet where they are initially adsorbed, thus preventing the formation of large Pt islands on the Ru nanoparticles, although Pt monomers may diffuse across some edges more efficiently than across others.
Of course, since the work reported here has not considered facets of the Ru nanoparticles that exhibit other geometries, the relative adsorption energies and diffusion barriers of these may expand and tune the occupancy landscape from the one we have inferred.

\subsubsection{Pt dimers} \label{prIIIB2}

The results of previous sections indicate that clustering of 2D Pt islands on the Ru facets occurs readily and that, although the Ru edges hinder the inter-facet diffusivity of Pt monomers, they do not unquestionably prevent it at room temperature. Yet, in order for Pt islands to coalesce and form larger islands, they would have to diffuse through the edges. We therefore turn to the calculation of the energy barrier for a dimer to diffuse from hcp to fcc sites (on the (0001)
facet) and through edge $B$, the \emph{easy} edge for monomers to cross.

Some of the sites that Pt dimers may adopt on the (0001) and the ($\overline{1}$101) facets, as well as the corresponding average formation energies per atom, are shown in Fig.~\ref{fig:pr7}.  We find that $E_{form}/n$ of the dimer increases (by $\sim$~0.12 eV) with respect to that of the monomer, suggesting that dimers would preferably form rather than diffuse as monomers through the \emph{easy} edges. As in the case of monomers, dimers prefer to sit on hcp sites on the (0001) facet (see Fig.~\ref{fig:pr7}(a) and (b)); similarly, the adsorption energy of Pt dimers at hcp sites near edge $B$ (only one edge-atom neighbor) and in the middle of the Ru stripe is almost the same (see Fig.~\ref{fig:pr7}(a) and (c)). On the (0001) facet, when one of the atoms in the dimer comes closer to the edge and its coordination is reduced from 5 to 4 (compare Fig.~\ref{fig:pr7}(c) and (d)), $E_{form}/n$ drops $\sim$~0.16 eV, suggesting that there is a higher barrier for Pt dimers to
approach the edges to the point where its atoms become more undercoordinated. On the ($\overline{1}$101) facet, analogously, $E_{form}/n$ is 0.52 eV lower for a dimer across the (0001) and ($\overline{1}$101) edges (see Fig.~\ref{fig:pr7}(e)) than for a dimer sitting on the ($\overline{1}$101) facet close to the edge (see Fig.~\ref{fig:pr7}(f)).

As shown in Fig.~\ref{fig:pr8}, the barrier for the dimer to diffuse from fcc to hcp sites has a height comparable to that for monomers while it is 2.5 times smaller for the inverse process, from hcp to fcc. The diffusion rates (1.4$\times$10$^9$ and 8.8$\times$10$^4 \: s^{-1}$, correspondingly) thus differ by four orders of magnitude, indicating that dimers are most of the time at hcp sites. This presents another indication that Pt dimers (or larger islands) would not leave the (0001) facet since, for it to diffuse across edge $B$, it must be on fcc sites.

As shown in Fig.~\ref{fig:pr9}, the diffusion across edge $B$ comprises two stages. The initial state of the first stage corresponds to the configuration shown in Fig.~\ref{fig:pr7}(d) in which one atom is on a hcp1 site and the other is slightly beyond the edge whereas, in the final state, the atom at the hcp1 site diffuses to a fcc4 site and the other moves to a type "2 " site (see Fig.~\ref{fig:pr5}(b)) of the ($\overline{1}$101) facet, as shown in Fig.~\ref{fig:pr7}(e). The energy barriers for the dimer to diffuse back and forth from the initial and final states are shown in Fig.~\ref{fig:pr9} while the corresponding diffusion rates for these processes are low ($\sim$~10$^4$). Indeed, the energy barriers of the first stage are very similar to those for dimer
diffusion from hcp to fcc sites (see Fig.~\ref{fig:pr8}) and for monomer diffusion through edge $A$ (see Fig.~\ref{fig:pr6}). Notice also that the final state described above is only an intermediate stage of the diffusion towards the ($\overline{1}$101) facet. For the second stage, the initial state is naturally the configuration shown in Fig.~\ref{fig:pr7}(e) while the final state corresponds to that shown in Fig.~\ref{fig:pr7}(f), in which both atoms sit on the ($\overline{1}$101) facet. The energy barrier to move from the initial to the final state (see Fig.~\ref{fig:pr9}) produces also a low diffusion rate of 6$\times$10$^3 \: s^{-1}$, while the inverse process, whose barrier is 3 times larger, provides a diffusion rate of  $\sim$1.4$\times$10$^{-14} \: s^{-1}$, indicating that dimers like monomers on the ($\overline{1}$101) facets will most likely remain there. In short, edge $B$, which offers a low-barrier diffusivity path to monomers, renders to Pt dimmers two energy barriers for facet-to-facet diffusion, each representing a diffusion rate at least three orders of magnitude lower than those of the monomer. We expect the diffusion rates of trimers and other n-mers of Pt on Ru nanoparticle facets to be even lower than that found here for dimers.

\section{Summary}\label{prIV}

We have calculated from first principles the energetics and geometry of Pt islands deposited on Ru(0001), as well as the energy barriers for diffusion of Pt monomers and dimers through the edges intersecting the (0001) and  ($\overline{1}$101) facets of a superstructure modeling a Ru nanoparticle. We find that the low coordination of Pt atoms composing the islands, the strong Pt-Ru interaction, and the hcp structure of the substrate promote formation of increasingly large Pt islands on Ru(0001), possibly up to 1ML, and avoid the 2D/3D crossover. On the other hand, the scenario is quite different when Pt atoms are deposited on Ru nanoparticles. In a simple model for (0001) factes with edges connecting to ($\overline{1}$101) orientations, we concur with experimental predictions that Pt atoms arrange homogeneously over the facets of Ru nanoparticles by spontaneous deposition and form 2D islands. We also predict that these islands do not coalesce into a large unique island because the edges of the Ru nanoparticles prevent monomers and dimers from diffusing to other facets. Our calculated barriers indicate that there may be some edges in the Ru nanoparticles for which the diffusion rate \emph{across-edge} is several orders of magnitude lower than the diffusion rate \emph{on-facet}, even for monomers. For those edges that may offer relatively low diffusion barriers to monomers, our calculated barriers for dimers suggest that dimers or larger islands, whose formation is more probable than the monomer diffusion, remain in the facet where they were formed since the diffusion rate \emph{across-edge} is several orders of magnitude lower than that of monomers.

\begin{acknowledgments}

We would like to thank R. Adzic for bringing this problem to our attention and for insightful discussions. This work was supported in part by US-DOE, under grant DE-FG02-07ER46354.

\end{acknowledgments}

\bibliography{ptru}

\clearpage

\begin{table*}
\caption[Adsorption energy of a Pt monomer on the Ru nanoparticle model]{\label{tab:tpr1}Adsorption energy (in eV) of a Pt monomer on various sites of the (0001) and ($\overline{1}$101) facets of our Ru nanoparticle model (see numbering notation in Fig.~\ref{fig:pr5})}
\begin{ruledtabular}
\begin{tabular}{|c|c|c|c|c|c|}
\multicolumn{2}{c}{0001 hcp}&\multicolumn{2}{c}{0001 fcc}&\multicolumn{2}{c}{($\overline{1}$101)}\\
\hline
Label  & $E_{ads}$  &Label  & $E_{ads}$  & Label & $E_{ads}$  \\
(Fig.~\ref{fig:pr5}) & (eV) & (Fig.~\ref{fig:pr5}) & (eV) & (Fig.~\ref{fig:pr5}) & (eV)\\
\hline
1 &     4.94&   4&      4.86&   1&      5.66 \\
\hline
2&      4.93&   5&      4.77&   2&      4.92 \\
\hline
3&      5.05&   6&      4.78& &         \\
\end{tabular}
\end{ruledtabular}
\end{table*}

\clearpage

FIG. 1. (color online) Model of the edges of a faceted Ru nanoparticle exposing a (0001) facet and two (\=1101) facets.
Different colors distinguish the five layers parallel to the (0001) surface constituting the
structure.

FIG. 2. Average formation energy per atom,
$E_{form}/n$, as a function of the size, $n$, of the island for
$n = $ 1 - 5 on the (3x4) supercell (upper panel) and for $n$ =1 - 4, 6, 7, and 9 on the (4x4)
supercell (lower panel).

FIG. 3. (color online) Two configurations of a 7-atom Pt island (blue) on Ru(0001) (grey)
showing the detachment of one Pt atom. The configuration in the
right panel has lower $E_{form}/n$ than the configuration
in the left panel by $\sim$0.14 eV (see text).

FIG. 4. (color online) Two configurations of a 9-atom Pt island (blue) with
2D(left) and 3D(right) structure on Ru(0001) (grey) used in calculations.

FIG. 5. (color online) Adsorption sites of Pt monomers (red)
on the (0001) facet of the Ru
nanoparticle model (blue). Numbers "1", "2" and "3" indicate  hcp sites and "4", "5",
"6" indicate fcc sites. (b) Adsorption sites of Pt monomers on the (\=1101) facets of the Ru
nanoparticle model. Numbers "1" and "2" indicate four-fold and
three-fold hollow sites, respectively.

\begin{figure}
\includegraphics[width=0.60\textwidth]{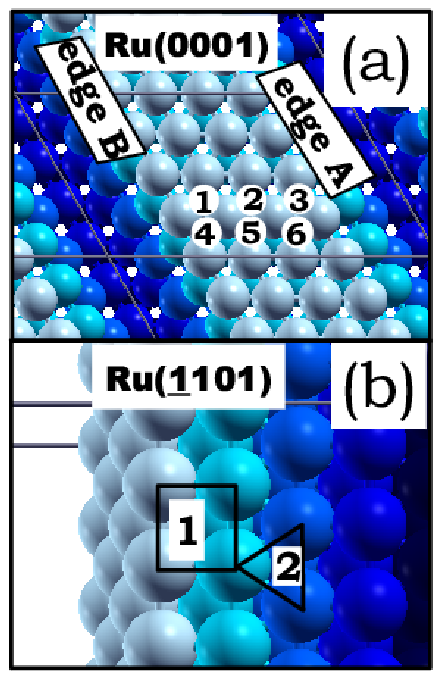}
\caption{\label{fig:pr5}}
\end{figure}

\begin{figure}
\includegraphics[angle=0, width=0.5\textwidth]{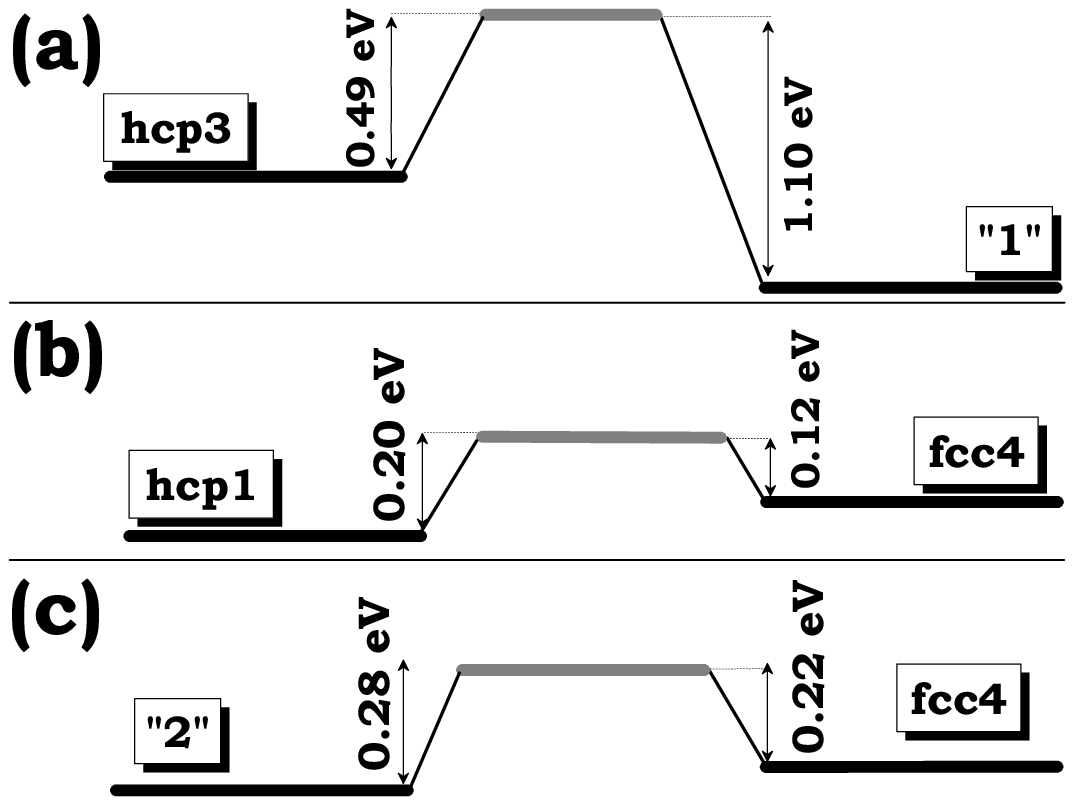}
\caption{\label{fig:pr6}}
\end{figure}

\begin{figure*}
\includegraphics[width=1.0\textwidth]{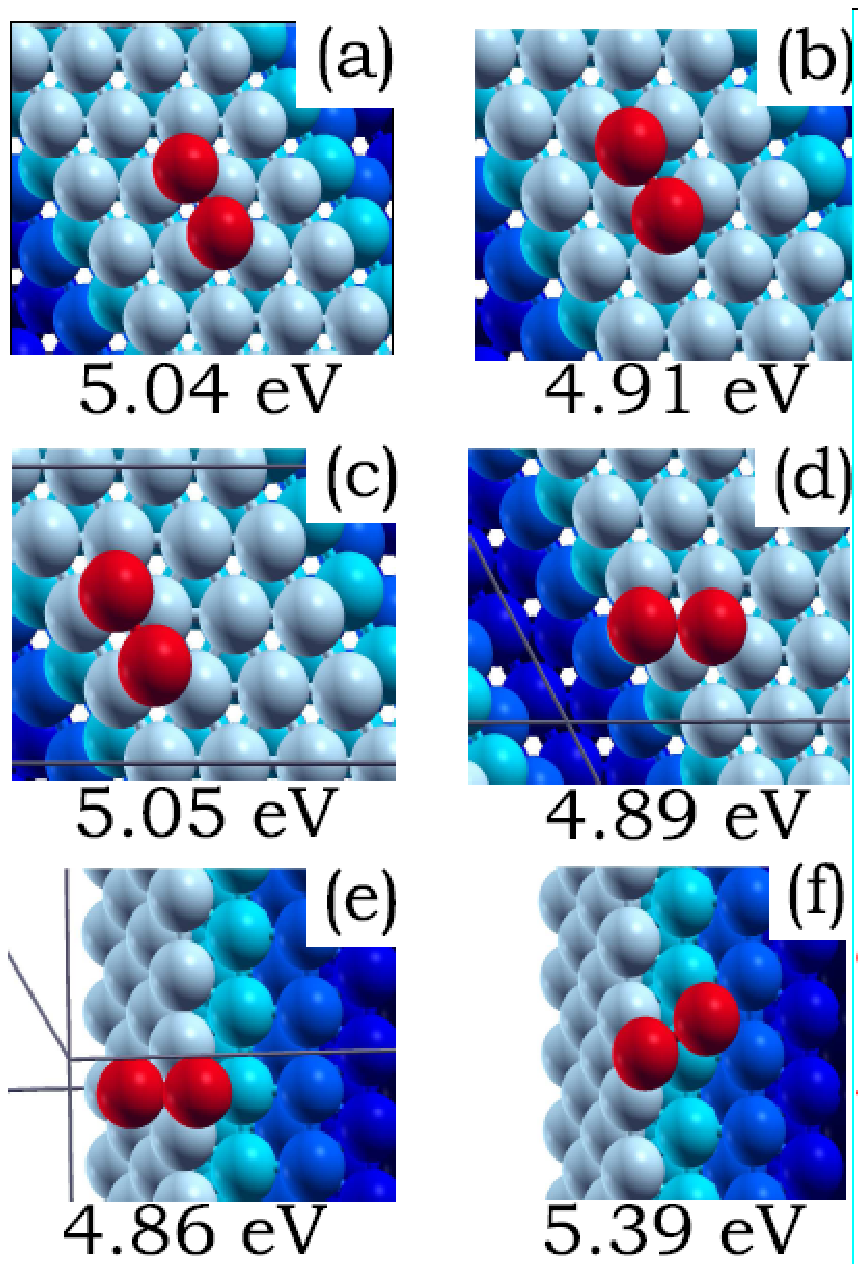}
\caption{\label{fig:pr7}}
\end{figure*}

\begin{figure*}
\includegraphics[width=1.0\textwidth]{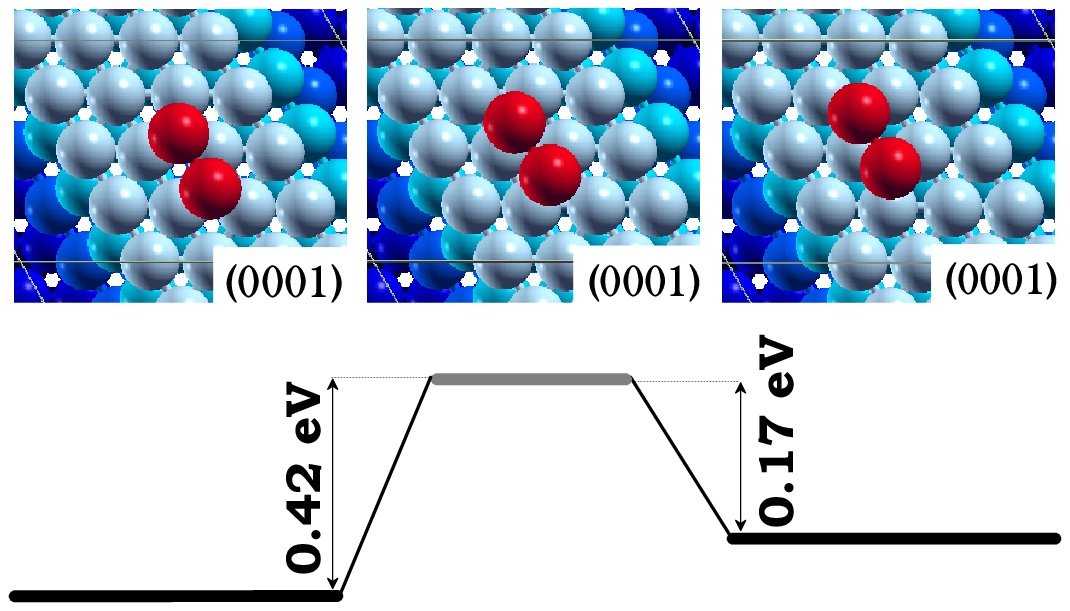}
\caption{\label{fig:pr8}}
\end{figure*}

\begin{figure*}
\includegraphics[angle=90, width=0.85\textwidth]{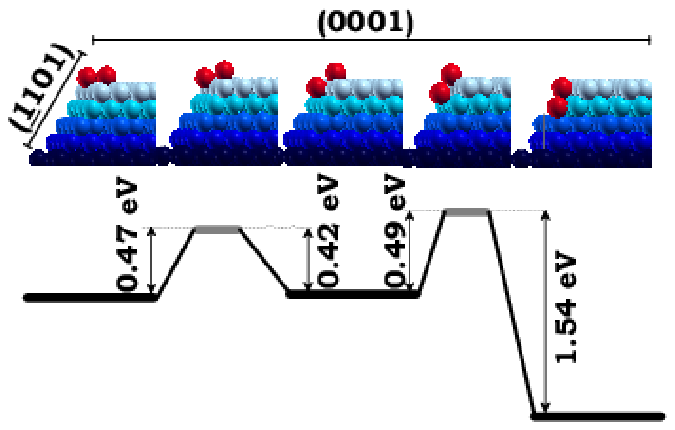}
\caption{\label{fig:pr9}}
\end{figure*}

FIG. 6. Energy barrier for the diffusion of a Pt monomer (a) across
the edge A (see Fig.~\ref{fig:pr5}): hcp3 $\longleftrightarrow$ type "1" site,
(b) on the (0001) facet: hcp1 $\longleftrightarrow$ fcc4, and (c)
across the edge B: fcc4 $\longleftrightarrow$ type "2" site.

FIG. 7. (color online) Adsorption sites and $E_{form}/n$ ($n$ = 2) of Pt dimers (red)
on the facets of the Ru nanoparticle model (blue). (a)  Top view of the (0001) facet
showing a dimer at hcp2 sites. (b) Top view of the (0001) facet
showing a dimer at fcc5 sites.
(c) Top view of the (0001) facet
showing a dimer at hcp1 sites.
(d) Top view of the (0001) facet
showing a dimer with one atom at a hcp1 site and the other beyond the edge
of the same facet.
The coordination number of the atoms is 5 and 4, respectively.
(e) Top view of the (\=1101) facet showing a dimer with one atom at a
fcc4 site and the other at a type "2"
hollow site. (f) Top view of the (\=1101) facet showing a dimer at
a type "2" and type "1" hollow sites of the same facet.

FIG. 8. (color online) The upper three panels from left to right
illustrate initial, transition, and final states, respectively, of the
diffusion of a dimmer (red) from hcp to  fcc sites, as seen from the top of
the (0001) facet (blue). The lower panel shows
the corresponding barriers of this process and the inverse, from fcc to hcp sites.

FIG. 9. (color online) The upper five panels from left to right
illustrate the two-step diffusion of the dimmer (red) across the edge
intersecting the (0001) and the (\=1101) facets (blue).
First, third, and fifth upper panels are local minimum energy
configurations of the dimmer and the second and forth upper panels are
transition states. The lower panel shows the barrier for the dimmer to
diffuse back and forth from either local minimum energy
configuration (see text).

\clearpage

\end{document}